\documentclass[twocolumn]{revtex4}
\usepackage{amsmath}
\usepackage{graphicx}
\usepackage{pifont}

\begin{document}
\title{The Quantum Interference Effect Transistor}
\author{D. M. Cardamone, C. A. Stafford, S. Mazumdar}
\affiliation{Department of Physics, University of Arizona, 1118 E. 4th Street,
  P. O. Box 210081, Tucson, AZ 85721}

\begin{abstract}
We propose a new type of molecular transistor, the Quantum Interference Effect
Transistor (QuIET), based on tunable current suppression due
to quantum interference. We show that any aromatic hydrocarbon ring has
two-lead configurations for which current at small voltages is suppressed
by destructive interference. A transistor can be created by providing phase
relaxation or decoherence at a site
on the ring. We propose several molecules which could tunably
introduce the necessary dephasing or decoherence, as well as a proof of principle using a
scanning tunneling microscope tip. Within the
self-consistent Hartree-Fock approximation, the QuIET is shown to have
characteristics strikingly similar to those of conventional field effect and
bipolar junction transistors.
\end{abstract}
\maketitle

\section{Introduction}
Although there has been considerable scientific and commercial interest, a
small molecular transistor has yet to be discovered. Needless to say, such a
device is crucial to transferring existing technology to smaller length
scales. Current industrial fabrication techniques have more or less exhausted
the possibilities for purely classical phenomena to solve this problem. We must
therefore turn to quantum mechanical effects in the search for smaller
transistors. The solution, therefore, is also of fundamental interest to
mesoscopic and molecular physicists: we find that the ``small transistor'' problem is an
engaging way of posing the question, ``What happens to multi-lead, many-body electronic
systems when their size is such that quantum effects are important?''

We propose a new type of device, the Quantum Interference Effect Transistor (QuIET), capable of filling the role
of the Field Effect Transistor (FET) and Bipolar Junction Transistor (BJT) at
length scales $\lesssim$1nm. The QuIET consists simply of a
hydrocarbon ring and a mechanism to tunably introduce
phase relaxation at a particular site. Unlike the Single Electron Transistor
(SET), 
the transistor behavior can occur over a large range of base voltages. Its
$I-\mathcal{V}$ characteristic is a single, broad resonance, strikingly similar to those of
macroscopic transistors over a domain of several volts. This is to be taken in
contrast to the SET's $I-\mathcal{V}$, a series of many sharp peaks.

\begin{figure}
\includegraphics[trim=2cm 1cm 2cm 2cm,keepaspectratio=true,width=\columnwidth,clip=true]{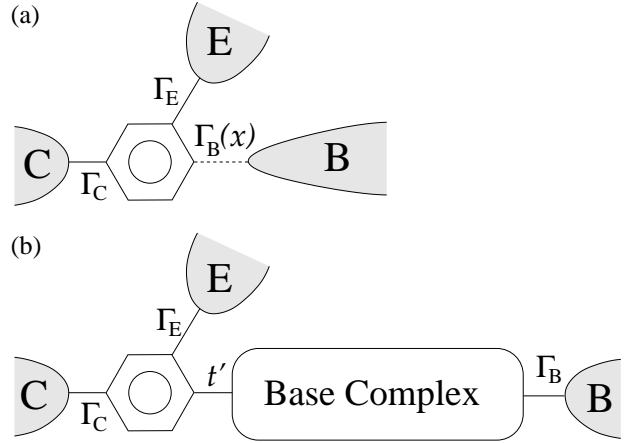}
\caption{Schematic diagrams of two types of QuIET. In each, base voltage modulates the coherent suppression of current
between emitter (E) and collector (C) leads. In (a), base voltage controls the
distance $x$ between the benzene ring and base lead (B), for example an STM tip. This in turn controls the coupling of the ring
to the base lead. In (b), a base complex is introduced between the
ring and base lead. The electrostatic effect of the base lead's bias on this molecule
alters its coupling to the benzene ring.}
\label{transistor}
\end{figure}

The operating principle of the QuIET is that bias applied to a third lead can
modulate an otherwise complete conductance suppression across a hydrocarbon
ring. This conductance suppression is a simple, single-particle effect of
quantum mechanics. The modulation from a third lead can be achieved either through
direct coupling to the
molecule (Fig.\ \ref{transistor}a), as in the case of a scanning tunneling
microscope (STM) tip, or via the introduction of an appropriate
intermediary molecular complex, as shown in Fig.\ \ref{transistor}b. In either case, the
resultant device is an excellent molecular-scale transistor.

Section \ref{model} gives an outline of the extended Hubbard model Hamiltonian
and Green function formalism we use to
treat the problem. The multi-terminal current formula allows extraction of
current. Section \ref{ring} explains the tunable coherent
conductance suppression by which the QuIET works. Section \ref{dba}
focuses on the use of a acceptor-donor molecule as a scalable method of
tuning interference. Section \ref{numerics} presents
numerical results which indicate the viability of this
idea. Finally, Section \ref{conclusion} contains our conclusions.

\section{Model}
\label{model}
The Hamiltonian of the system can be written as the sum of three terms:
$H=H_m+H_l+H_{tun}$. The first is the extended Hubbard model
molecular Hamiltonian
\begin{equation}
\label{Hm}
H=\sum_{i\sigma}\varepsilon_i d_{i\sigma}^\dagger
d_{i\sigma}+\sum_{\langle ij\rangle\sigma}t_{ij}\left(d_{i\sigma}^\dagger
  d_{j\sigma}+\mathrm{H.c.}\right)+\sum_{ij}\frac{U_{ij}}{2}Q_iQ_j,
\end{equation}
where $d_{i\sigma}$ annihilates an electron on atomic site $i$ with spin
$\sigma$, $\varepsilon_i$ are the site energies, and $t_{ij}$ are the tunneling
matrix elements. The final term of Eq.\ (\ref{Hm}) contains intersite
and same-site Coulomb interactions, as well as the electrostatic effects of the
leads. The interaction energies are  modeled according to the Ohno
parameterization \cite{ohno}:
\begin{equation}
\label{interactions}
U_{ij}=\frac{11.13\mathrm{eV}}{\sqrt{1+.6117\left(R_{ij}/\mathrm{\AA}\right)^2}},
\end{equation}
where $R_{ij}$ is the distance between sites $i$ and $j$.
$Q_i$ is an effective charge operator for atomic site $i$:
\begin{equation}
Q_i=\sum_\sigma d_{i\sigma}^\dagger d_{i\sigma}-\sum_\alpha\frac{C_{i\alpha}\mathcal{V}_\alpha}{e}-1.
\end{equation}
The second term represents the polarization charge on site $i$ due to
capacitive coupling with lead $\alpha$. Here $C_{i\alpha}$ is the capacitance between site $i$ and lead $\alpha$, chosen to correspond with the interaction energies of Eq.\
(\ref{interactions}), and $\mathcal{V}_\alpha$ is the voltage on lead
$\alpha$. $e$ is the magnitude of the electron charge.

The QuIET is intended for use at room temperature and above, a temperature range far beyond the regime in
which lead-molecule or lead-lead correlations play an important role. As such,
we have followed the method of Ref.\ \cite{capacitance} and treated electrostatic
interactions between molecule and leads at the level of capacitance parameters. The electronic situation of the
leads is thus completely determined by the externally controlled
voltages $\mathcal{V}_\alpha$, along with the leads' temperatures and Fermi
energies. Each lead possesses a continuum of states, and their total Hamiltonian is
\begin{equation}
\label{Hl}
H_l=\sum_{\alpha}\sum_{\substack{k\in\alpha\\ \sigma}}\varepsilon'_kc_{k\sigma}^\dagger
c_{k\sigma},
\end{equation}
where $\varepsilon'_k$ are the energies of the single-particle levels $k$ in
lead $\alpha$, and $c_{k\sigma}$ are the annihilation
operators for the states in the leads.

Tunneling between molecule and leads is provided by the final term of the
Hamiltonian:
\begin{equation}
\label{Htun}
H_{tun}=\sum_{\langle
  i\alpha\rangle}\sum_{\substack{k\in\alpha\\ \sigma}}\left(V_{ik}d_{i\sigma}^\dagger
  c_{k\sigma}+\mathrm{H.c.}\right).
\end{equation}
$V_{ik}$ are the tunneling matrix elements for moving from a level $k$ within
lead $\alpha$ to the nearby site $i$. Coupling of the leads to the ring via
inert molecular chains, as may be desirable for fabrication purposes, can be
included in the effective $V_{ik}$, as can the effect of the substituents
used to bond the leads to the molecule.

A system whose Hamiltonian includes such terms as Eqs.\ (\ref{Hl}) and
(\ref{Htun}) requires an
infinite-dimensional Fock space, but we
are concerned mainly with the behavior of the discrete molecule
suspended between the leads. We therefore adopt a Green function approach, in which
Dyson's Equation gives the Green function of the full system
\begin{equation}
\label{Dyson}
G(E)=\left[G_m^{-1}(E)-\Sigma(E)\right]^{-1},
\end{equation}
where $G_m$ is the Green function of the isolated molecular system. With the
use of an appropriate self-energy $\Sigma$, Equation
(\ref{Dyson}) is true both for the retarded Green function $G^r$ as well as for its
$2\times 2$ Keldysh counterpart.

The retarded
self-energy due to the leads is
\begin{equation}
\Sigma_{ij}^r(E)=-\frac{i}{2}\Gamma_i(E)\delta_{ij},
\end{equation}
where the energy widths are given by Fermi's Golden Rule
\begin{equation}
\label{FGR}
\Gamma_i(E)=2\pi\sum_\alpha\sum_{k\in\alpha}|V_{ik}|^2\delta\left(E-\varepsilon'_{
    k}\right).
\end{equation}

We take the broad-band limit of Eq.\ (\ref{FGR}) and treat each
of the $\Gamma_i$ as a constant parameter characterizing the lead-site coupling. The
only effect of $H_l$ and $H_{tun}$ in this limit is to shift the poles of the
Green function into the complex plane. This causes
the density of states
$\rho(E)=-\frac{1}{\pi}\mathrm{Im}\mathrm{Tr}G^r(E)$ to change from a discrete spectrum
of delta functions
to a continuous, width-broadened function. Due to the open nature of the
system, electrons can occupy all energies.

The retarded Green function gained via Eq.\ (\ref{Dyson}) contains all
information regarding the dynamics of the system. In particular, the current in lead $\beta$ is given by the familiar multi-terminal current
formula \cite{buettiker}:
\begin{equation}
\label{l-b}
I_\beta=\frac{2e}{h}\sum_\alpha\int_{-\infty}^\infty
dE\;T_{\alpha\beta}(E)\left[f_\alpha(E)-f_\beta(E)\right],
\end{equation}
where $f_\alpha$ is the Fermi function for lead $\alpha$. The transmission
probability is
\begin{equation}
T_{\alpha\beta}(E)=\Gamma_a\Gamma_b|G_{ab}^r(E)|^2.
\end{equation}
Here $a$($b$) is the site with hopping to lead
$\alpha$($\beta$). We note that Eq.\ (\ref{l-b}) is an exact result of the
Keldysh formalism in cases, like ours, consisting only of elastic processes.

In order to arrive at $G^r(E)$, we must consider electron-electron interactions. Here we do so via the well known self-consistent Hartree-Fock
method. The Hamiltonian is replaced by its mean-field approximation
\begin{widetext}
\begin{equation}
\label{hfham}
H_m^\mathrm{HF}=\sum_{i\sigma}\left(\varepsilon_i-\sum_{j\alpha}U_{ij}\frac{C_{j\alpha}\mathcal{V}_\alpha}{e}\right)d_{i\sigma}^\dagger
d_{i\sigma}+\sum_{\langle ij\rangle\sigma}t_{ij}\left(d_{i\sigma}^\dagger
  d_{j\sigma}+\mathrm{H.c.}\right)
+\sum_{ij\sigma\rho}U_{ij}\left(\langle
  d_{j\rho}^\dagger d_{j\rho}\rangle d_{i\sigma}^\dagger
  d_{i\sigma}-\langle d_{j\rho}^\dagger d_{i\sigma}\rangle
  d_{i\sigma}^\dagger d_{j\rho}\delta_{\sigma\rho}\right).
\end{equation}
\end{widetext}
In this approximation, the retarded Green function is $G^r_m(E)=\left(E-H^\mathrm{HF}_m+i0^+\right)^{-1}$.

Equation (\ref{hfham}) gives the mean-field Hamiltonian as a function of the
diagonal and off-diagonal equal-time correlation functions $\langle
d_{i\sigma}^\dagger d_{j\sigma}\rangle$. To complete a self-consistent loop,
we require an expression for these quantities in terms of Green functions. They
are given in the Keldysh formalism by the equal-time limit of the ``$<$''
Green function
\begin{equation}
G_{i\sigma,j\sigma}^<(t,t')=i\langle d_{i\sigma}^\dagger
(t)d_{j\sigma}(t')\rangle=\int_{-\infty}^\infty\frac{d\omega}{2\pi}G_{i\sigma,j\sigma}^<(\omega)\mathrm{e}^{-i\omega(t-t')}.
\end{equation}

From Dyson's Equation (\ref{Dyson}) it follows \cite{keldysh} that
\begin{equation}
G^<=G^r\Sigma^<G^{r\dagger}+(1+G^r\Sigma^r)G_m^<(1+\Sigma^{r\dagger}G^{r\dagger}).
\end{equation}
The second term is a purely equilibrium property, and can be related to the
total charge on the molecule when the three lead biases are equal.
The ``$<$'' self-energy is given by
\begin{equation}
\Sigma^<_{ab}(\omega)=i\Gamma_af_\alpha(\omega)\delta_{ab}.
\end{equation}
The desired relation between the equal-time
correlation functions and $G^r$,
\begin{equation}
\langle d_{i\sigma}^\dagger
d_{j\sigma}\rangle=\sum_a\Gamma_a\int_{-\infty}^\infty\frac{d\omega}{2\pi}G^r_{i\sigma,a\sigma}(\omega)G^{r*}_{a\sigma,j\sigma}(\omega)f_\alpha(\omega),
\end{equation}
is now readily computed, and the self-consistent loop is complete.

\section{Tunable conductance suppression}
\label{ring}
\begin{figure}
\begin{center}
\includegraphics[width=\columnwidth,keepaspectratio=true]{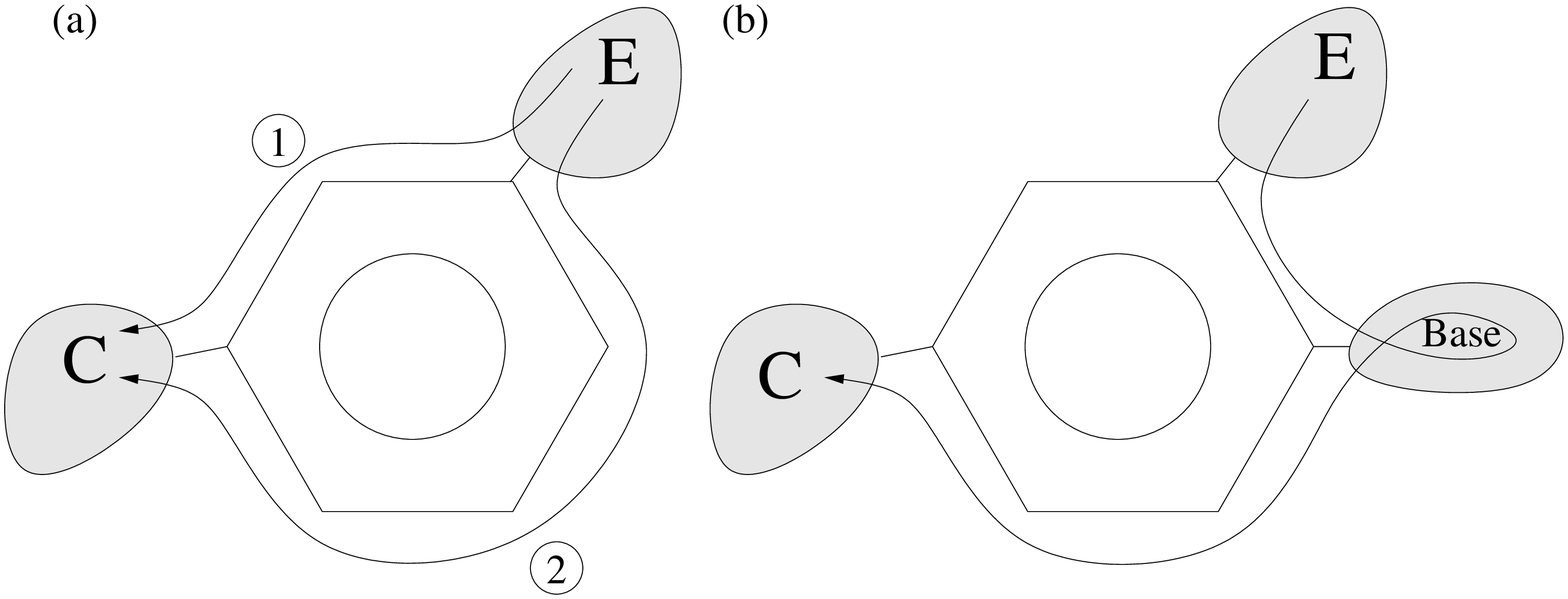}
\end{center}
\caption{(a) Two-lead experiment to measure the conductance of benzene when
  the leads are in the meta configuration. Shown are the two most direct paths
  a carrier can take from the emitter lead to the collector. These two cancel
  exactly, as do all other paths with the same endpoints
  in a similar pairwise fashion. (b) Example of a new path allowed
  when a base complex or lead is included. Such paths are not
  canceled, and so contribute to the total current between emitter and collector.}
\label{2paths}
\end{figure}
In the two-lead device shown in Fig.\ \ref{2paths}a, a single carrier
(electron or hole) is injected into the ring by the
emitter and exits via the collector some time later. In the path
integral formulation of quantum mechanics, it traverses all paths around the
ring allowed by the connectivity of the system during this process.

We operate the QuIET in the regime where there is
little charge transfer between it and the leads. In the
linear response,
 the carrier has momentum equal to the Fermi momentum of the ring
$k_F=\frac{\pi}{2a}$, where $a=1.397\mathrm{\AA}$ is the intersite spacing of
benzene. Clearly, then, the phase difference between paths \ding{172} and
\ding{173} is $\pi$, and they cancel exactly. Similarly, all of the paths through the ring from emitter to collector
exactly cancel in a pairwise fashion. Therefore, transport of carriers is
forbidden in linear response. 

\begin{figure*}
\setlength\unitlength{.333\textwidth}
\begin{picture}(3,1)
\put(0,0){\includegraphics[width=.65\columnwidth,keepaspectratio=true]{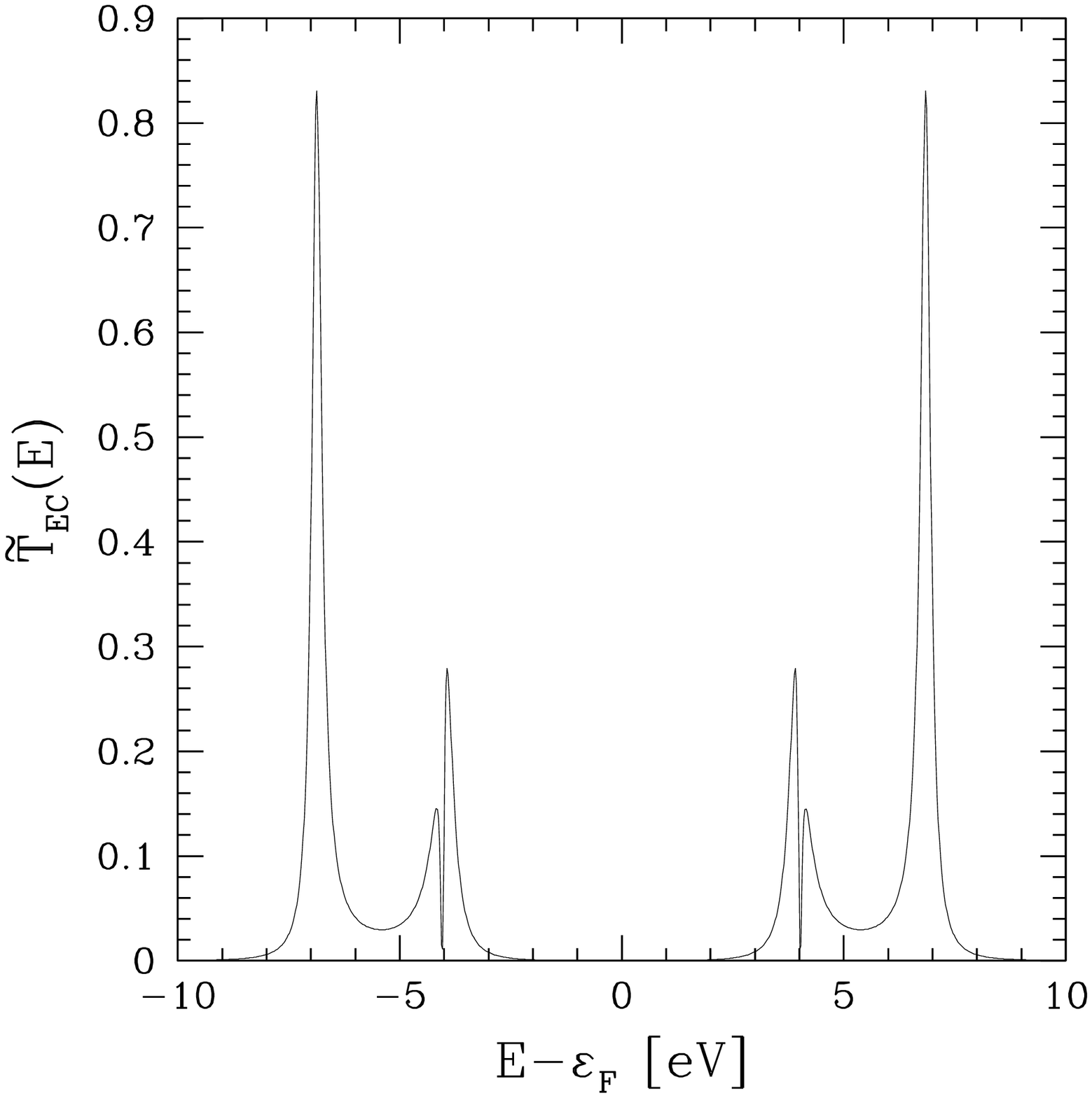}}
\put(1,0){\includegraphics[width=.65\columnwidth,keepaspectratio=true]{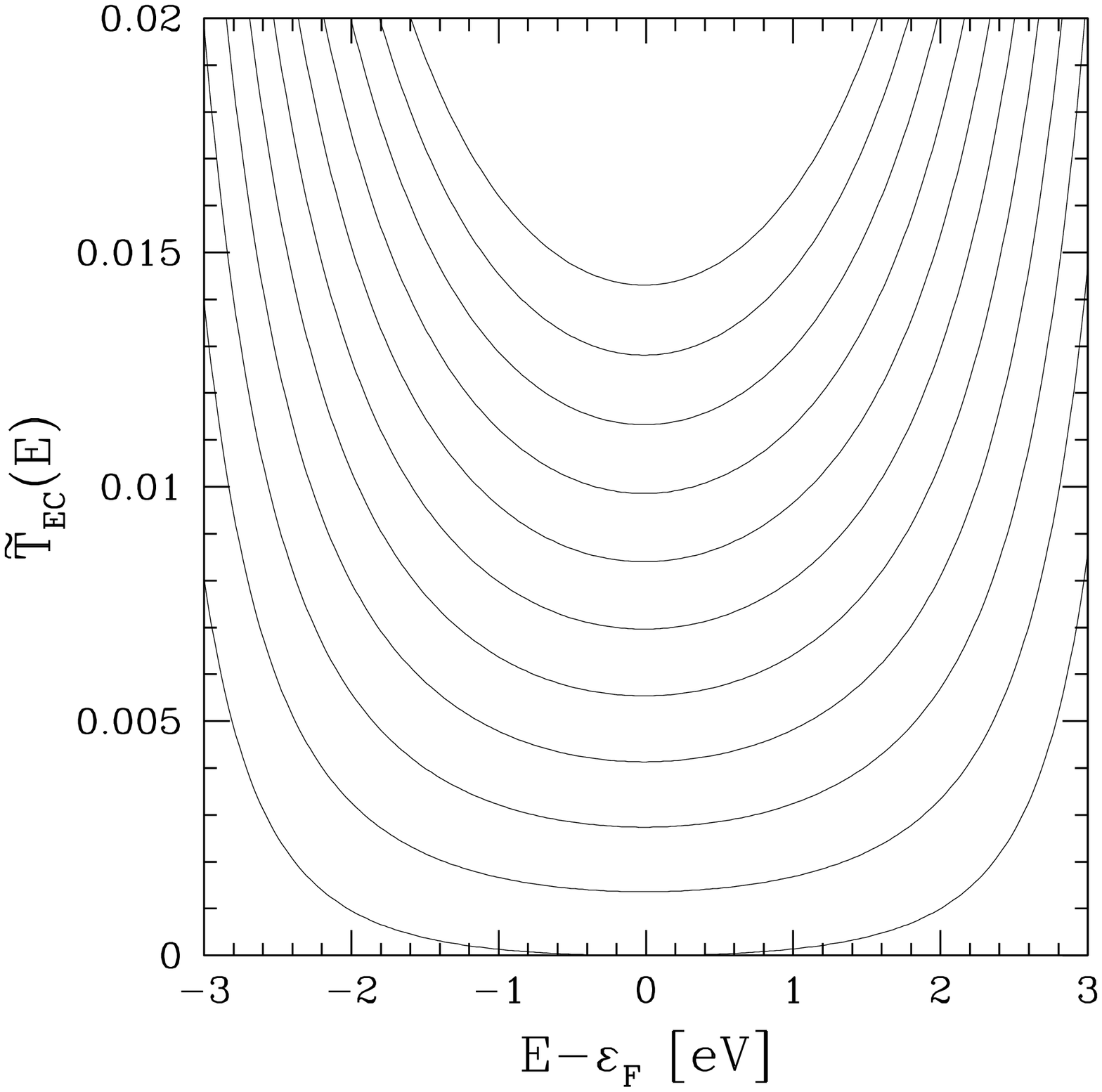}}
\put(2,0){\includegraphics[width=.65\columnwidth,keepaspectratio=true]{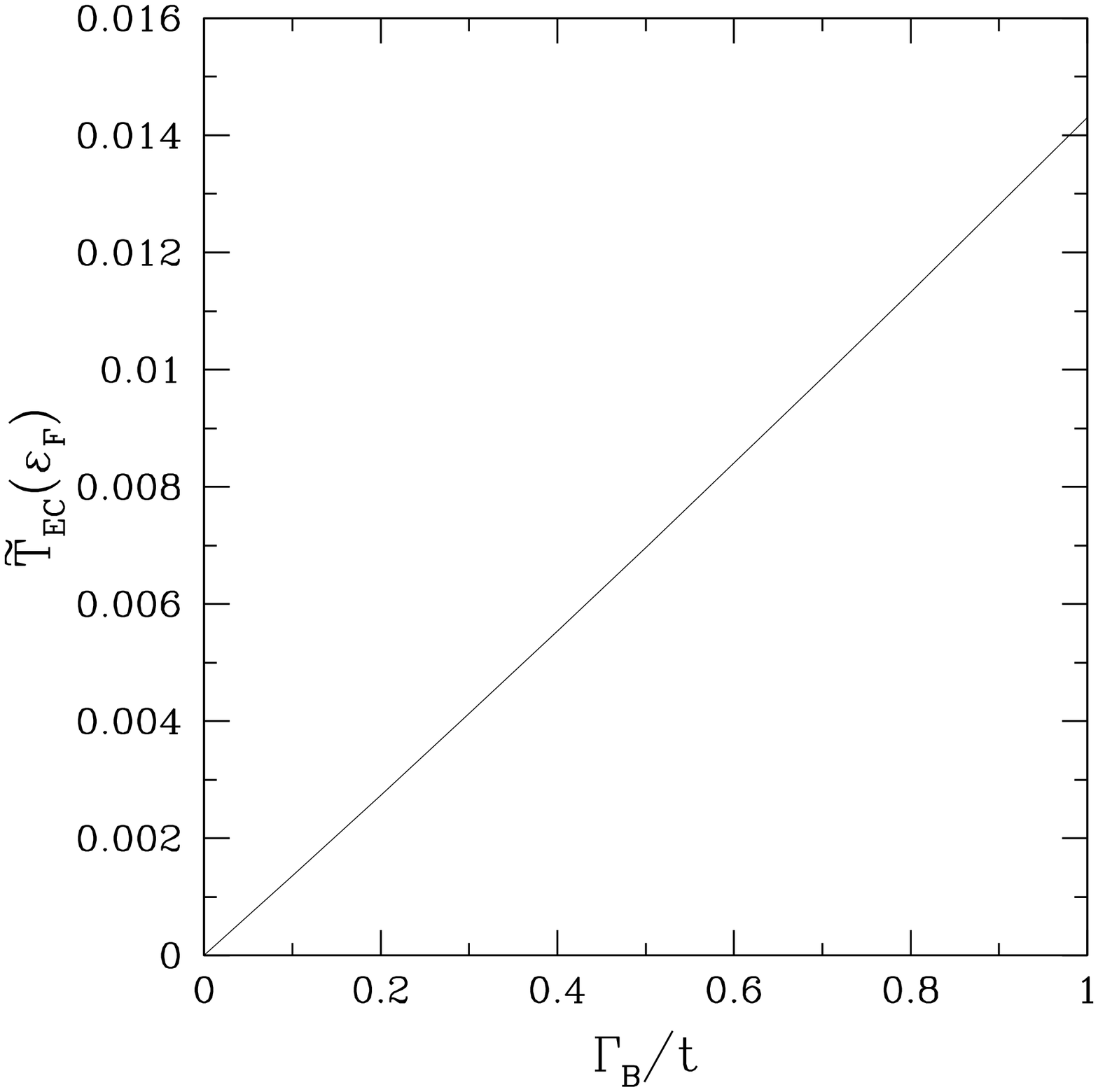}}
\put(0,.85){(a)}
\put(.95,.85){(b)}
\put(1.95,.85){(c)}
\end{picture}
\caption{Total transmission probability of the device
  shown in Fig.\
  \ref{transistor}a. $\tilde{T}_{EC}=T_{EC}+\frac{T_{EB}T_{BC}}{T_{EB}+T_{BC}}$,
  where the second term is due to a restriction that the base lead be an
  infinite-impedance ideal voltage
  probe \protect\cite{datta}. Here, $\Gamma_E=.5t$ and $\Gamma_C=.2t$, but the choice of energy
  widths does not affect the qualitative results. In (a), $\Gamma_B=0$. In (b), the lowest curve shows $\Gamma_B=0$ and each
  successive curve shows increasing $\Gamma_B$ in increments of $.1t$. (c) shows
  the transmission at the Fermi energy as a function of $\Gamma_B$.}
\label{suppression}
\end{figure*}

It is a consequence of Luttinger's Theorem \cite{luttinger} that this coherent suppression of current persists into the interacting
regime, as demonstrated in Fig.\ \ref{suppression}a. The transmission is calculated by the self-consistent Hartree-Fock model outlined in
Section \ref{model}. In this figure the base coupling $\Gamma_B=0$, and so
transmission at the Fermi energy is wholly suppressed by coherence. Figure
\ref{suppression}b shows how this result changes as $\Gamma_B$ is
increased. Current is allowed to flow due to two new phenomena. The first is
that new paths, such as the one shown in Fig.\
\ref{2paths}b, are added. These paths have no particular phase relationship to
other paths, and so support transmission. Additionally, nonzero $\Gamma_B$
denotes decoherence, and therefore a departure from the perfect coherent
current suppression of $\Gamma_B=0$. This is the basic operating principle of
the QuIET: coherent current suppression can be tunably broken by the
introduction of decoherence and dephasing from a base complex or third lead. In fact, Figure \ref{suppression}c shows that
the transmission varies nearly linearly for $\Gamma_B\le t$.

Clearly, the transistor behavior based on the mechanism outlined above is requisite upon an
assumption that the device be operated well within the gap of
benzene. The numerical simulations discussed in Section \ref{numerics} indicate that the best transistor results are
found for collector-emitter bias $\lesssim 1-2\mathrm{V}$. Another, related,
consideration is that in equilibrium, charge transfer between the molecule and
leads not play an important role. For this to be true, the work function of
the metallic leads must be comparable to the chemical potential of
benzene. Fortunately, this is the case with many bulk metals, among them palladium,
iridium, platinum, and gold \cite{crc}.

\begin{figure}
\includegraphics[keepaspectratio=true,width=\columnwidth]{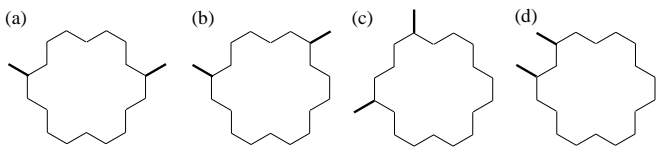}
\caption{Emitter-collector lead configurations possible in a QuIET
  based on [18]-annulene. The bold lines represent the positioning of the two
  leads. Each of the four arrangements has a different phase difference
  associated with it: (a) $\pi$, (b) $3\pi$, (c) $5\pi$, and (d) $7\pi$.}
\label{18}
\end{figure}

While in this work we have chosen to focus on benzene, the QuIET mechanism
applies to any aromatic annulene with leads positioned so the two most
direct paths have a phase difference of $\pi$. Furthermore,
larger molecules have other possible lead configurations, based on phase
differences of $3\pi$, $5\pi$, etc. Figure \ref{18} shows the lead
configurations for a QuIET based on [18]-annulene. Of course, benzene
provides the smallest of all possible QuIETs.

\begin{figure}
\includegraphics[keepaspectratio=true,width=\columnwidth]{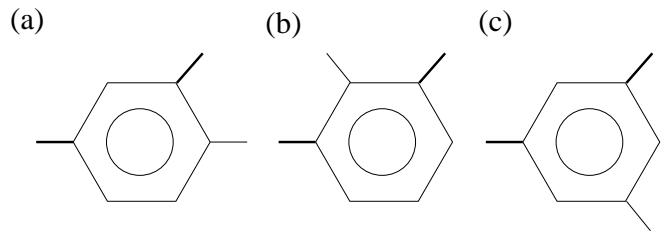}
\caption{The three different arrangements for a third lead on
  benzene when the emitter and collector leads (bold) are in the meta
  configuration. (a) is the choice which couples most strongly to the
  conducting orbitals of benzene. (b) is a second case which allows for QuIET
  operation. The third possibility, (c), decouples entirely from the
  conducting molecular orbitals by symmetry. A third lead in this configuration cannot break the coherent current suppression at
  all, and so the molecule does not function as a QuIET.}
\label{3rdlead}
\end{figure}

The position of the third lead affects the degree to
which destructive interference is suppressed. For benzene, the most effective
location for a third lead is shown in Fig.\ \ref{3rdlead}a. The base may also be
placed at the site immediately between the emitter and collector leads, as
shown in Fig.\ \ref{3rdlead}b. The
QuIET operates in this configuration as well, although since base coupling to
the current carriers is less, the transistor effect is somewhat
suppressed. The third, three-fold symmetric configuration of leads (Fig.\
\ref{3rdlead}c) completely
decouples base from current carriers within benzene. Because of this, the base
cannot be used to provide the decoherence or dephasing necessary to QuIET
operation in this configuration. For each aromatic hydrocarbon, exactly one
three-fold symmetric lead configuration exists and yields no transistor behavior.

\section{Base Complex}
\label{dba}
With the tunable current suppression outlined in the previous section, the
working principle of a transistor has become apparent. The most straightforward method of
varying $\Gamma_B$ is to
change the distance of the third lead from the hydrocarbon molecule, as via an
STM tip. It is also
possible, however, to interpose an additional molecular complex between
annulene and the base lead. If the effect of biasing the base lead is to increase
the transparency of the molecular complex, dephasing and decoherence result. Such a
QuIET is depicted schematically in Fig.\ \ref{transistor}b.

\begin{figure}
\includegraphics[trim=3cm 1cm 9cm 2cm,keepaspectratio=true,width=\columnwidth]{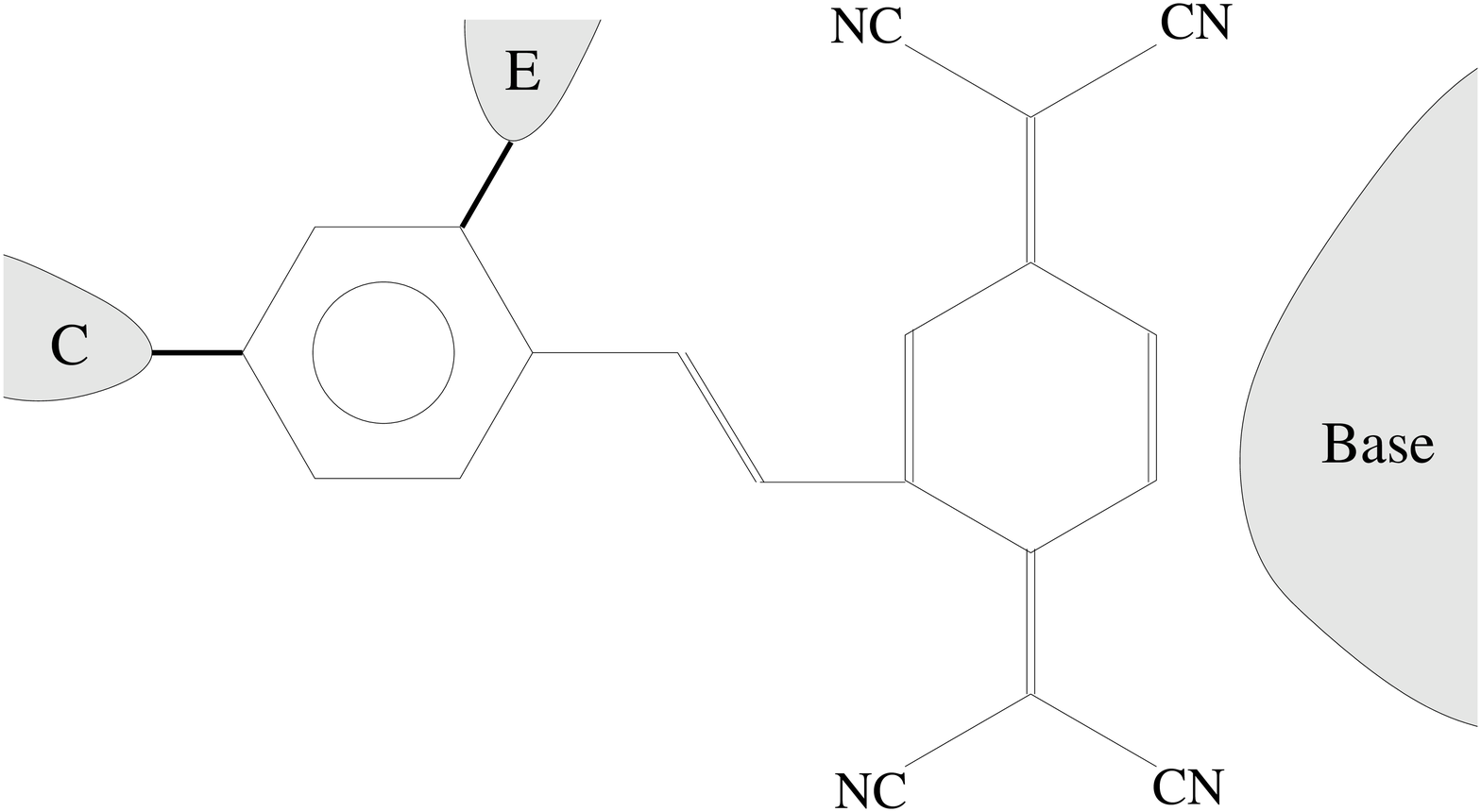}
\caption{Specific example of a QuIET: a dithiol derivative of
  phenylene--TCNQ. E and C are  the emitter and collector leads. The
  base lead can be either weakly coupled via TCNQ, as for a BJT analogue, or
  capacitively coupled if FET-like behavior is desired.}
\label{QuIET}
\end{figure}

A simple way to engineer the base
complex is to include two spatially separated orbitals, with
a detuning from each other $\Delta$ and hopping to each other $t_B$. If $\Delta\gg t_B$ for
zero bias on the base lead, paths through the base complex are highly
suppressed. Since the orbitals are spatially separate, however, one
electrostatically couples more strongly to the base lead. Thus, the voltage on
that lead can be used to control $\Delta$, and hence change the transparency
of the base complex. The result is the broad conductance peak characteristic
of two discrete levels coming into resonance.

One way to achieve this structure is simply to link the annulene via an inert
bridge to a donor or acceptor molecule, \emph{e.\ g.} the dithiol derivative of
phenylene--TNCQ shown in Fig.\ \ref{QuIET}. In this case, the effect of a base
bias of appropriate sign is to bring the lowest unoccupied molecular orbital (LUMO) of TCNQ into resonance with the
neighboring $\pi$ orbital of phenylene. The extra paths allowed within and
through the TCNQ break the coherent suppression of current
within the benzene ring. Any donor or acceptor molecule with which phenylene
bonds in this way will yield similar results, although a donor used in this
way will allow emitter-collector current to flow for negative, rather than
positive, base voltages.

The class of donor-acceptor rectifiers proposed by Aviram and Ratner
\cite{aviram74} have this two-level structure as well. These molecules possess 
an orbital of increased electron affinity (the acceptor) and one of
increased ionization potential (the donor), as well as perhaps an inert bridge
orbital. By the influence of a base lead, the donor highest occupied molecular
orbital (HOMO) and acceptor LUMO can be brought into resonance, and the
rectifier be made to serve as an effective QuIET base complex. In addition to
this effect, several many-body mechanisms have been proposed whereby
such molecules could exhibit asymmetric current flow \cite{waldeck93}. For our
purposes, however, it is sufficient to note that rectification in general is
indicative of a tunable transparency to charge carriers. 

The variety within this well studied
family of molecules (\emph{e.~g.}, Refs.\ \cite{aviram74,waldeck93,geddes90,metzger97,metzger03,heath03}) lends
a great deal of versatility to the QuIET. Different choices of donor,
acceptor, and bridge molecules allow QuIETs to be fabricated for specific
applications. Table \ref{specs} gives approximate ranges of parameters
available for the use of molecular diodes as base complexes.

\begin{table}
\caption{Approximate ranges of values available for molecular diodes which could be used
  as QuIET base complexes.}
\label{specs}
\begin{tabular}{|l|r|}
\hline
Parameter & Range\\
\hline
\hline
Donor HOMO energy & [-8,-5]eV\\
\hline
Acceptor LUMO energy & [-3,-1]eV\\
\hline
$t_B$ & [.01,1]eV\\
\hline
$t'$ & [.01,1]eV\\
\hline
Distance between donor and acceptor & [6,10]\AA\\
\hline
\end{tabular}
\end{table}

\section{Numerical Results and Discussion}
\label{numerics}
We turn now to discussion of numerical results, based on the self-consistent
Hartree-Fock
model presented in Section \ref{model}. For the purposes of illustration, the
numerical results are based on the dithiol derivative of phenylene--TCNQ shown
in Fig.\ \ref{QuIET} and three bulk gold leads. Similar results can be
obtained for many such QuIETs.

For the purposes of electrostatic interactions, each lead is considered to be immediately adjacent to its neighboring site. The electrostatic interactions between TCNQ
and phenylene are modeled as those between two conducting spheres of
appropriate size, while within phenylene
the Ohno parameterization (\ref{interactions}) is used. We neglect
reduction of the $U_{ij}$ due to screening from a third site, an effect which
we find to be of the order of 1\%. In keeping with our assumption that the
leads do not significantly perturb the molecular system, modifications to the
interaction parameters due
to the presence of the leads are also neglected.

\begin{figure}
\setlength\unitlength{\columnwidth}
\begin{picture}(1,2)
\put(0,1){\includegraphics[width=\columnwidth,keepaspectratio=true]{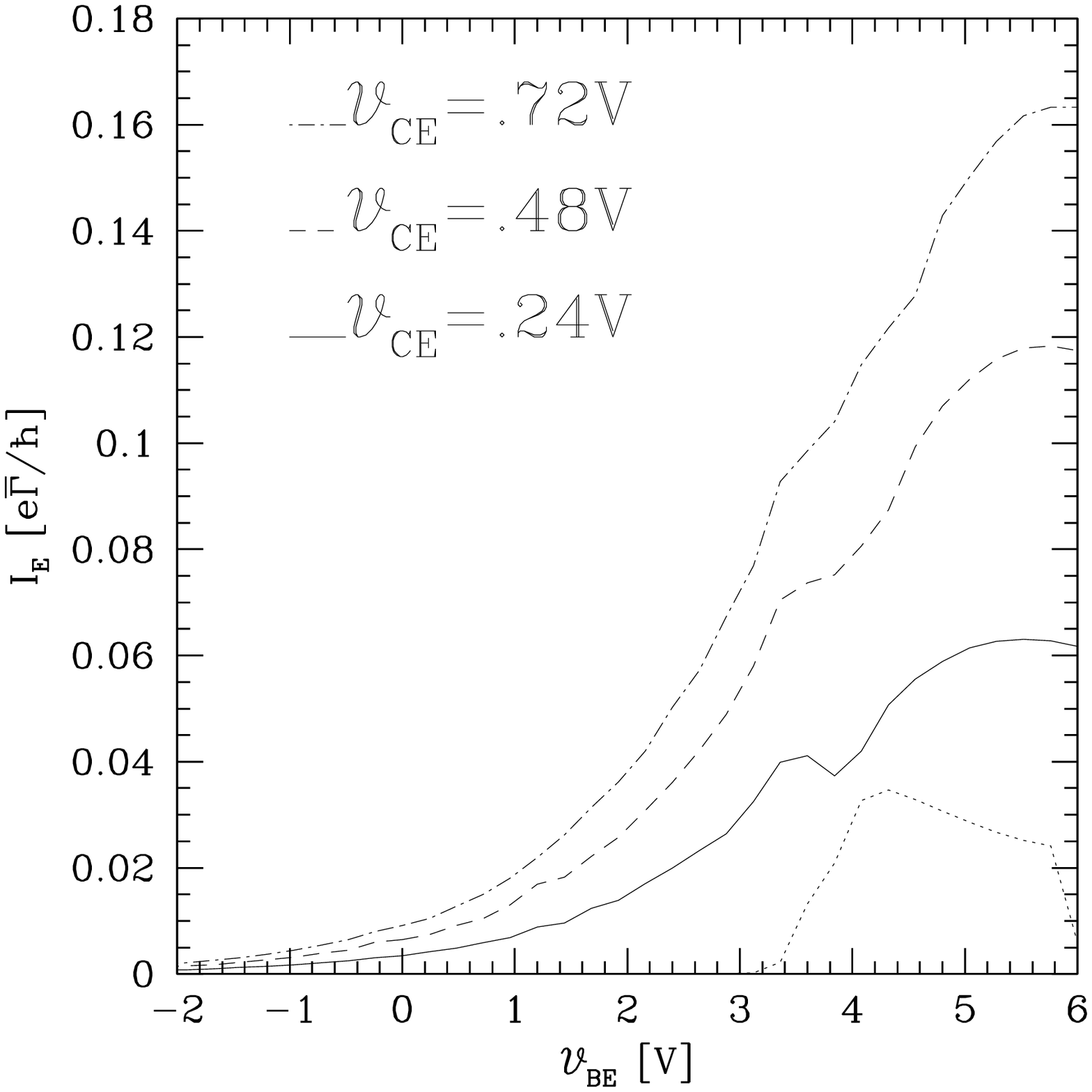}}
\put(0,0){\includegraphics[width=\columnwidth,keepaspectratio=true]{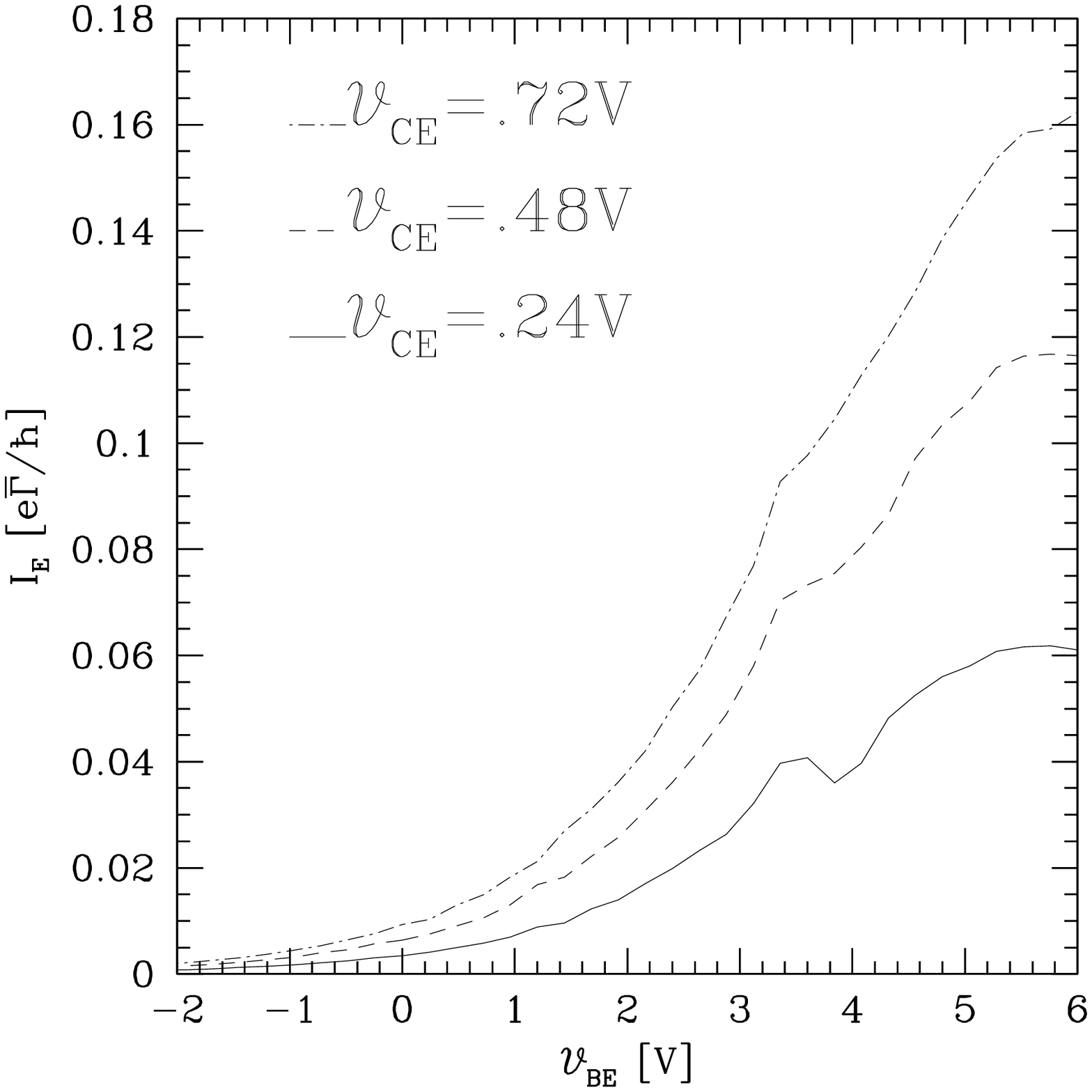}}
\put(-.01,1.9){(a)}
\put(-.01,.9){(b)}
\end{picture}
\caption{Typical $I-\mathcal{V}$ characteristic of two QuIETs, showing
  emitter current out of the molecule
  vs.\ voltage on the base lead
  for three different biases applied to the
  collector, with the emitter ground. The calculation is done for the molecule
  in Fig.\ \ref{QuIET} at room
  temperature.
  $\overline{\Gamma}\equiv\frac{\Gamma_E\Gamma_C}{\Gamma_E+\Gamma_C}$ gives the
  sequential tunneling rate through the device. Here,
  $\Gamma_E=\Gamma_C=2.4\mathrm{eV}$. In (a), $\Gamma_B=.0024\mathrm{eV}$, and the QuIET amplifies
  current, similar to a BJT. The dotted curve is the base current into the molecule for the case
  of $\mathcal{V}_{CE}=.24\mathrm{V}$, multiplied by 10 for clarity. Base currents for other collector voltages are
  similar. (b) shows the FET limit of the QuIET, with $\Gamma_B=0$.}
\label{IV}
\end{figure}
A typical $I-\mathcal{V}$
diagram for this QuIET is shown in Fig.\ \ref{IV}, demonstrating that
the QuIET is quite
reminiscent in operation to classical transistors. The currents in
the emitter and collector leads exhibit a broad resonance as the
base voltage is increased. Furthermore, for nonzero $\Gamma_B$, the device amplifies the current
in the base lead, providing emulation of the classical BJT. Since a donor,
TCNQ, was used in the base complex, the device performs in a manner analogous
to an NPN transistor, with base current flowing into the molecule. Use of an
acceptor instead, for example TTF, yields PNP-like behavior. For capacitive
coupling of the third lead,
on the other hand,
a FET-like $I-\mathcal{V}$ characteristic is obtained.

We interpret this transistor behavior as due to the coherence
mechanism discussed in Section \ref{ring}. If hopping between the benzene ring
and the base complex is set to zero, we find that full coherent
current suppression is restored and almost no current flows between the
emitter and collector
leads. Furthermore, the current step effect persists for arbitrarily small 
$\Gamma_B$, which is consistent with our interpretation that transport
through the
non-canceling paths is enhanced by the electrostatic effect of the third
lead.

Estimation of the $\Gamma_i$'s remains an open question in the field of
molecular electronics. Similar quantities are often estimated to be
$\lesssim .5\mathrm{eV}$ \cite{nitzan01,paulsson03} by the method of Ref.\
\cite{mujica94}, whereas values as high as 1eV have been suggested
\cite{tian98}. For the emitter and collector broadenings, we have taken values
somewhat higher than the norm for the sake of numerical
convergence. Fortunately, nothing in the arguments of Sections \ref{ring} or
\ref{dba} depends strongly on the choice of these quantities. So long as they
are positive, their magnitude merely determines the scale of the current, and
has negligible effect on the overall operating principle and transistor
behavior of the QuIET.

\begin{figure}
\includegraphics[width=.9\columnwidth,keepaspectratio=true]{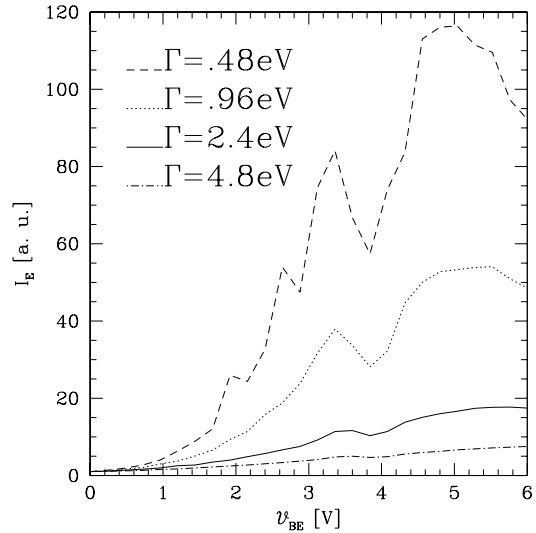}
\caption{$I_E$ vs. $\mathcal{V}_{BC}$ for different values of
  $\Gamma_E=\Gamma_C=\Gamma$. Each curve has been scaled so that the leakage
  current at zero bias is 1. $\Gamma_B=0$,
  $\mathcal{V}_{CE}=.24\mathrm{V}$, and positive current is
  out of the molecule. Reducing $\Gamma$ suppresses the leakage current, but
  does not strongly affect the QuIET's behavior on-resonance, thus greatly
  enhancing its overall function.}
\label{gammabar}
\end{figure}

Figure \ref{gammabar} demonstrates the effect of varying the coupling of the
benzene ring to emitter and collector leads. As is to be expected, the internal
structure of the resonance, due to electrons moving to screen the changing
field from the base, sharpens. Furthermore, at low base voltages, less than
about .75V, the base complex is strongly off resonance, and the scale of the
leakage 
current through the annulene is set mainly by
$\overline{\Gamma}\equiv\frac{\Gamma_E\Gamma_C}{\Gamma_E+\Gamma_C}$, an expected
result for systems like the QuIET, which are not near a charge fluctuation resonance \cite{stafford96}.

For higher base voltages, the base complex begins to play an important
role. Thus, as described by Ref.\ \cite{paulsson03}, $\overline{\Gamma}$
alone no longer determines the scale of the current. Instead, we find that the
rate-limiting process is travel through the base complex, and varying
$\Gamma_E$ and $\Gamma_C$ has little effect. Thus, smaller emitter and
collector energy widths enhance the QuIET's current contrast dramatically.

\section{Conclusions}
\label{conclusion}
We have presented a novel idea for a small single-molecule transistor. The device is based on the coherent
suppression of current through an aromatic hydrocarbon ring, and the control of that effect
by increasing the contribution of paths outside the ring. Two simple methods
of creating that control, either through an STM tip or through a small base
complex, were discussed, and numerical results presented.

In contrast to
SET-style devices, the QuIET is predicted to be an extremely versatile,
scalable device. Since it is chemically constructed, it should be possible to
fabricate as many identical QuIETs as desired. Furthermore, QuIETs can be
designed which mimic the functionality of all major classes of macroscale
transistors: FET, NPN, and PNP.

\section*{Acknowledgments}
The authors acknowledge support from National Science Foundation Grant Nos.\ PHY0210750,
DMR0312028, and DMR 0406604.

\end{document}